\documentclass[aps,ams,amsmath,prx,twocolumn,superscriptaddress]{revtex4-2}
\usepackage{graphics}
\usepackage{graphicx}
\usepackage{amsmath}
\usepackage{amsfonts}
\usepackage{bm}
\usepackage{color}
\usepackage{bbm}
\usepackage{hyperref}
\def\bm#1{\mbox{\boldmath{$#1$}}}

\newcommand{\be}{\begin{equation}}
\newcommand{\ee}{\end{equation}}

\hypersetup{colorlinks=true,linkcolor=blue,anchorcolor=blue,citecolor=blue,filecolor=blue,urlcolor=blue,bookmarksnumbered=true,pdfview=FitB}

\usepackage{bm}
\usepackage{color}
\usepackage{xcolor}
\usepackage{bbm}
\usepackage[normalem]{ulem}
\usepackage{comment}
\usepackage{soul}
\usepackage[varg]{txfonts}

\newcommand{\beq}{\begin{equation}}
\newcommand{\eeq}{\end{equation}}
\newcommand{\beqa}{\begin{eqnarray}}
\newcommand{\eeqa}{\end{eqnarray}}
\newcommand{\bea}{\begin{eqnarray}}
\newcommand{\eea}{\end{eqnarray}}
\usepackage{amssymb}
\usepackage{array}
\usepackage{amsmath}
\usepackage{graphicx}\usepackage{graphics}
\usepackage{dcolumn}
\usepackage{bm}\usepackage{varioref}

\begin{document}

\title{
Axion electrodynamics and giant magnetic birefringence in Weyl excitonic insulators
}

\author{A. A. Grigoreva}

\affiliation{\hbox{Department of Physics, University of Washington, Seattle, Washington 98195, USA}}

\author{A. V. Andreev}

\affiliation{\hbox{Department of Physics, University of Washington, Seattle, Washington 98195, USA}}

\author{L. I. Glazman} 
\affiliation{\hbox{Department of Physics and Yale Quantum Institute, Yale University, New Haven, Connecticut 06520, USA}}

\date{\today}

\begin{abstract}
We study the electromagnetic (EM) response of the excitonic insulator phase of a time-reversal (TR) invariant Weyl semimetal (WSM). At low temperatures, the system develops two exciton condensates. The condensates are related to each other by TR symmetry and weakly coupled by a Josephson tunneling term. The latter leads to the formation of the Leggett mode~\cite{Leggett} with a small gap. Our main finding is that the Leggett mode couples to the EM fields as a massive dynamical axion. This is a consequence of the chiral anomaly and the chiral magnetic effect in the parent WSM. Due to the small axion mass, its coupling to EM fields results in a giant anisotropic polarizability and birefringence in the presence of a static magnetic field. The photon-axion hybridization produces a polariton resonance near the axion gap.
\end{abstract}

\maketitle
 
At low temperatures, compensated semimetals may undergo a transition to a correlated insulating state, wherein electrons and holes become bound into neutral excitons~\cite{Keldysh,CLOIZEAUX,Jerome}, which form a quantum-coherent condensate. This causes reconstruction of the electron spectrum producing a gap at the Fermi level. The spectrum of the realigned bands is determined by the self-consistent exciton pairing potential. Therefore, excitonic dielectrics host collective modes of exciton condensate, which are not present in band insulators~\cite{Kozlov,Jerome, Guseinov, Keldysh_rev}. 
 
The topology of the Berry connection of the electron bands in the semimetal has a profound effect on the excitonic phase.
In Weyl semimetals, the \emph{s}-wave excitonic pairing with a fully gapped spectrum is only possible between electrons and holes from Weyl nodes with opposite chirality~\cite{Haldane}. Even in this case, the surface remains conducting, making the system a topological insulator~\cite{Grigoreva}. For brevity, we refer to the excitonic phase of WSMs as Weyl excitonic insulators (WEIs).

Here, we show that in WEIs that are invariant under time-reversal (TR) symmetry, the phase $\vartheta$ of the exciton condensate
couples to the electromagnetic field as a massive dynamical axion -- the hypothetical particle whose existence was proposed by Peccei and Quinn~\cite{peccei1977some} to solve the strong CP problem in quantum chromodynamics. The coupling of axions to the electric and magnetic fields $\bm{E}$ and $\bm{B}$ is described by the following term in the Lagrangian density,
\begin{align}
    \label{eq:axion_coupling_intro}
    \mathcal{L}^{\mathrm{ax}} =  & \,   \frac{e^2}{2\pi h c}\,  \bm{E} \cdot \bm{B}\, \vartheta ,
\end{align}
where $e$ is the electron charge, $c$ is the speed of light, and $h=2\pi \hbar$ is the Planck's constant. The axion coupling leads to unusual electrodynamics
\cite{wilczek1987two}. As is clear from Eq.~\eqref{eq:axion_coupling_intro}, the axion field $\vartheta$ is odd under TR. 
Previously,  realizations of axion electrodynamics have been proposed in systems with broken TR symmetry~\cite{essin2009magnetoelectric,li2010dynamical,qi2011topological,wang2013chiral,nenno2020axion}. The axion coupling of the exciton condensate in WEIs is in stark contrast with the situation in conventional excitonic insulators~\cite{Jerome,Zittartz,Halperin-Rice}, where the collective modes of the exciton condensate do not couple to the electric field. It also differs from the situation in TR-invariant axion band insulators \cite{essin2009magnetoelectric,qi2011topological}, where the axion field is static and is determined by the topology of the filled electron bands. In such cases, the axion coupling produces only surface effects. 
In contrast, the dynamical axion coupling significantly affects the dielectric response and propagation of electromagnetic waves in WEIs. The origin of the axion coupling of the phase of the exciton condensate to the electromagnetic field can be traced to the chiral anomaly~\cite{Bell, Adler} and chiral magnetic effect (CME)~\cite{Vilenkin} in the electron bands of the parent WSM. 

To simplify our arguments, we consider the minimal model of a TR-invariant WSM with four Weyl nodes. From TR symmetry it follows that the pairs of nodes of the same chirality must be located at opposite momenta.   Without loss of generality, we assume that the two positive chirality nodes located at momenta $\pm \bm{K}$ are electron-doped, and the two negative chirality nodes located at $\pm \bm{K}'$ are hole-doped. In a compensated semimetal, the densities of electrons and holes must be the same. We assume that the separation between the Weyl nodes exceeds the Fermi momentum, $k_F\ll |\bm{K} \pm  \bm{K}'|$,  and we also assume weak coupling so that the exciton binding radius is large compared to $k_F^{-1}$. 
To leading order in  $k_F/|\bm{K} \pm  \bm{K}'|$, the inter-valley transitions caused by electron-electron (\emph{e-e}) interactions may be neglected. In this approximation, the exciton correlations between electrons and holes in different nodes may be described by the long-wavelength Hamiltonian of the form 
\begin{align}
    \label{eq:H_long_wavelength}
    \hat{H}_0 = & \,  \sum_{a,{\bf k}} \xi_a ({\bf k}) c^\dagger_a({\bf k})c_a ({\bf k}) + \sum_{ab, |{\bf q}|< q_0} \frac{V_{ab} ({\bf q})}{2 \mathcal{V}} \hat{n}_a ({\bf q}) \hat{n}_b (-{\bf q}) .
\end{align}
Here $\mathcal{V}$ is the volume of the system, the indices $a$ and $b$ label the Weyl nodes, $c_a ({\bf k})$ is the electron annihilation operator, the density operators are given by $\hat{n}_a  ({\bf q}) = \sum_{{\bf k}}c^\dagger_a({\bf k})c_a ({\bf k}+{\bf q})$, and momenta ${\bf k}$ are measured from node $a$. 
At low energies and weak coupling, only the electrons with energies $\xi_a ({\bf k}) $ near the Fermi level are relevant. Therefore only the electron momenta below the cutoff $q_0 \sim k_F$ are included. 
In systems with congruent electron- and hole-Fermi surfaces, the excitonic instability develops for arbitrary weak repulsive interaction~\cite{Keldysh}. 
In our model of TR-invariant WSM, this situation is realized for isotropic dispersion
$\xi_a ({\bf k}) = \eta_a v_a \left(|{\bf k}| - k_F \right) $, where $v_a$ is the node velocity and $\eta_a=\pm 1$ is the node chirality. At weak coupling, the excitonic phase of the semimetal may be described by the mean-field
decoupling of the interaction~\cite{Keldysh,CLOIZEAUX,Jerome}
via the excitonic order parameter as
\begin{align}
    \label{eq:anomalous_average_new}
     \langle c_a^\dagger ({\bf p}) c_b ({\bf p}) \rangle  = & \,  e^{i \vartheta_{ab}} \psi_{ab} ({\bf p}).
\end{align}
Here $\psi_{ab} ({\bf p})$ is the exciton ``wavefunction"~\cite{CLOIZEAUX} normalized to the condensate density, and $\vartheta_{ab}$ is the condensate phase. The mean-field decoupling of the interaction causes reconstruction of the single-electron bands, creating a gap $\Delta$ at the Fermi level.

\begin{figure}[h!]
    \centering
\includegraphics[width=0.45\textwidth]{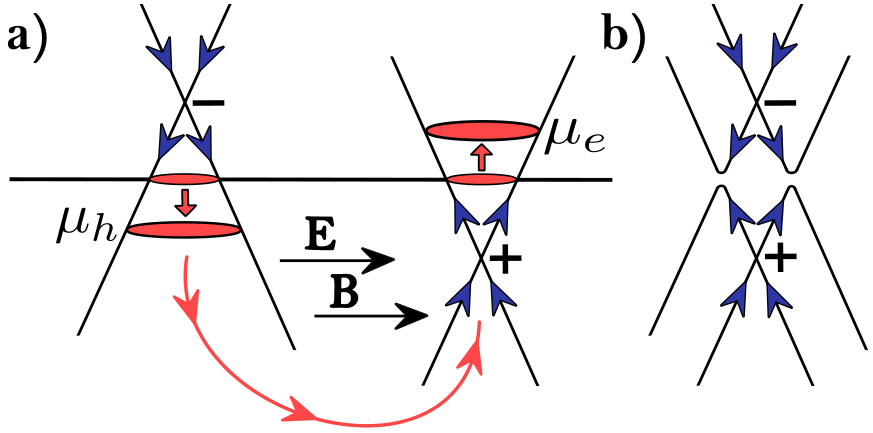}
    \caption{a) Fermi surface (FS) evolution caused by the spectral flow associated with the chiral anomaly, $\propto \bm{E}\cdot \bm{B}$. The equilibrium FSs (small ellipses) evolve to new positions shown by the large ellipses. For an isotropic spectrum, the FSs remain congruent, and the system remains unstable to exciton pairing for $\mu_h \neq \mu_e$. b) Reconstructed spectrum and Berry curvature flux (shown by blue arrows). 
    } 
  \label{fig:ChiralFlow}
\end{figure}

Let us show that in the approximation of the Hamiltonian \eqref{eq:H_long_wavelength}, the excitonic state of a Weyl semimetal is not an insulator in the presence of an external magnetic field $\bm{B}$, despite the presence of a gap in the single-particle spectrum. This is a consequence of the topological nature of WSM and can be understood as follows. Because of the chiral anomaly~\cite{NIELSEN_NLMR}, an application of a longitudinal electric field $\bm{E}\parallel \bm{B}$ to the system creates a flux of electron states (the spectral flow) between the valence and conduction bands across the Weyl nodes. The rate of the electron density change in valley $a$, caused by the spectral flow is  $\dot{n}_a=k_a \frac{e^2}{h^2c}\bm{E}\cdot \bm{B}$, where $k_a=\pm 1$ is the chirality of the node~\footnote{This rate can be understood as follows~{\color{magenta}\cite{NIELSEN_NLMR}}. The spectral flow arises from the zeroth Landau band, which has a chiral spectrum $\epsilon  (p_z) = k_a v p_z$, where $p_z$ is the momentum along $\bm{B}$. The acceleration of the electrons by the electric field, $\dot{p}_z = e E_z $, creates the flow of electron levels (spectral flow) between the valence and conduction bands. The spectral flow rate is given by the product of the Landau level degeneracy $e B/{hc}$ and $ \dot{p}_z/h$.}. Since the electron-doped and hole-doped nodes have opposite chirality, the densities of electrons and holes change at the same rate, and the electron- and hole-Fermi surfaces remain congruent, as shown in Fig. \ref{fig:ChiralFlow}. As a result, the exciton pairing and the gap in the single-electron spectrum are preserved by the spectral flow. On the other hand, the long-wavelength Hamiltonian in Eq.~\eqref{eq:H_long_wavelength} conserves the number of electrons in each valley. Therefore, the electron-electron interaction leading to excitonic pairing cannot relax the change of electron density in a given valley induced by the spectral flow. Thus, for $\bm{E}\cdot \bm{B} \neq 0$, the system cannot be in a stationary insulating state. 

The spectral flow creates an imbalance between the chemical potentials of the electron-filled valleys, $\mu_e$, and hole-filled valleys, $\mu_h$. This, in turn, creates a persistent current along $\bm{B}$, which is caused by the chiral magnetic effect (CME)~\cite{Vilenkin},
\begin{align}
    \label{eq:j_CME_def}
    \bm{j}_{\mathrm{CME}} = &  \, \frac{e^2}{h^2c} \bm{B} (\mu_e -\mu_h).
\end{align}
The chemical potentials are functions of the carrier density in each valley. Since in the absence of inter-valley \emph{e-e} scattering the latter are conserved, the excitonic state of a Weyl semimetal in the approximation of the Hamiltonian \eqref{eq:H_long_wavelength} is an ideal conductor capable of supporting a persistent current parallel to $ \bm{B}$~\footnote{In a finite system, this would produce a charge buildup at the boundary, creating macroscopic electric fields and longitudinal oscillations, see also Eq.~\eqref{eq:omega_L} below.}.

This capability allows the exciton condensate in WEIs to couple to the electric field, in stark contrast to the conventional (non-topological) excitonic insulators where the exciton condensate does not affect the electromagnetic response in the long-wavelength limit. Therefore, in a conventional excitonic insulator, the response is adequately described by the single-particle picture with realigned bands. In WEIs at $\bm{B}\neq 0$, the electric response must involve the exciton condensate. As we show below, the phase mode of the exciton condensate couples to the electromagnetic field
as a dynamical axion.

To obtain a quantitative description of the axion coupling, we note that since the exciton condensation occurs at the Fermi level, the condensate phase $\vartheta_{ab}$ changes at the rate $\hbar \dot{\vartheta}_{ab}= \mu_e-\mu_h$~\footnote{This is a condition that the exciton condensation opens a spectral gap at the local chemical potential. It applies not only in equilibrium but also in the presence of external perturbations with frequencies $\hbar \omega \ll \Delta$}. Thus, the density of CME persistent current in the excitonic state can be expressed in the form 
\begin{align}
    \label{eq:j_CME_phase}
    \bm{j}^{\mathrm{CME}}_{ab} = & \,  \frac{e^2 }{2 \pi h c} \, \bm{B} \,  \dot{\vartheta}_{ab}.
\end{align}
For each pair of the Weyl nodes connected by exciton pairing, the Lagrangian density, $\delta \mathcal{L} = \bm{j}_{\mathrm{CME}} \cdot \bm{A}/c$, describing the coupling of $\bm{j}_{\mathrm{CME}}$ to the vector potential $\bm{A}$, may be transformed to the form of   the axion coupling in Eq.~\eqref{eq:axion_coupling_intro}. This can be done by using
 $\bm{E}=-\dot{\bm{A}}/c$ and subtracting a total derivative, which does not affect the equations of motion of the dynamic axion field~\cite{Landau2}.

\emph{Low energy description:}
For energies below $\Delta$, the single-particle excitations and the amplitude mode of the exciton condensate may be ignored, and the effective action for the system may be expressed in terms of the condensate phases only. To be specific, we assume that the strongest excitonic instability corresponds to the pairing between the nodes with momenta $ \bm{K} $ and $ \bm{K}'$, and also $- \bm{K} $ and $ -\bm{K}'$, as required by the TR symmetry. We denote the phases of the corresponding excitonic condensates by $\vartheta_\alpha$, where
$\alpha=\pm $ labels the two pairs of Weyl nodes coupled by the exciton pairing.  The low-energy Lagrangian density can be written in  the form
\begin{align}
    \label{eq:L_0_CDW}
    \mathcal{L}^{(0)} \!= \! & \! \sum_\alpha  \left[ \frac{\nu  \left(\hbar \dot{\vartheta}_\alpha \right)^2 }{2  } \!-\frac{\rho  (\hbar \bm{\nabla} \vartheta_\alpha)^2   }{2 } \! -
    \alpha   \dot{\vartheta}_\alpha \bm{V}\cdot \bm{\nabla}\vartheta_\alpha +    \vartheta_\alpha  \frac{e^2 \bm{B}\cdot\bm{E}}{ 2\pi h c}    \right] .
\end{align}
Here, 
the first term, with $\nu$ being the thermodynamic density of states for each condensate, accounts for the ``kinetic" energy density induced by the inter-valley chemical potential imbalance, and the second term (with $\rho$ being the ``superfluid" density of the exciton condensate) describes the  ``potential" energy associated with the gradients of the condensate phases. The pair of nodes associated with condensate $\alpha$ can be characterized by a vector $ \alpha \bm{V}$, which is odd under time-reversal. This allows for a linear in $\bm{\nabla} \vartheta_\alpha$ term in Eq.~\eqref{eq:L_0_CDW}.

The variation of the Lagrangian \eqref{eq:L_0_CDW} with respect to the electric field yields  the electrical polarization $\bm{P} =\delta \mathcal{L}/\delta \bm{E}$ caused by the axion coupling,
\begin{align}
    \label{eq:Polarization_E}
    \bm{P} = & \, \frac{e^2 \bm{B}}{2\pi hc} \sum_\alpha \vartheta_\alpha. 
\end{align}
The polarization induced by an \emph{ac} electric field of the form $\bm{E}= \mathrm{Re} \left( \bm{E}_\omega e^{ - i \omega t} \right)$ is obtained by solving the equations of motion for $\vartheta_\alpha$ that follow from the Lagrangian \eqref{eq:L_0_CDW}. Relating it to the density of induced \emph{ac} current by $\bm{j}=\dot{\bm{P}}$ we obtain the frequency-dependent conductivity tensor induced by the axion coupling,
\begin{align}
    \label{eq:conductivity_ideal}
    \sigma^{(0)}_{ij} (\omega) = & \frac{2i}{\nu \hbar^2} \left( \frac{e^2}{2\pi h c} \right)^2 \frac{B_iB_j}{\omega}.
\end{align} 
The frequency dependence of the \emph{ac} conductivity here corresponds to a purely inductive response of an ideal conductor with the electron density $\propto B^2$. Alternatively, this conductivity may be expressed in terms of the effective plasmon frequency, $\sigma (\omega) = i\Tilde{\omega}_p^2/4\pi \omega $, where $\Tilde{\omega}_p^2 \sim  \omega_p^2/(k_F l_B)^4$. Here $l_B=\sqrt{\hbar c/eB}$ is the magnetic length, and $\omega_p$ is the plasma frequency  of the WSM, 
$\omega_p^2 \sim 4\pi e^2 k_F \epsilon_F/\hbar^2$.

Let us now account for the inter-valley transitions induced by the electron-electron interactions. Momentum conservation implies that to leading order in the small parameter $k_F/|{\bf K} \pm {\bf K}'|$ such processes are described by Hamiltonian of the form
\begin{align}
    \label{eq:H_pert}
    \!\hat{H}_1 \! = & \!  \sum_{{\bf p},{\bf k},{\bf q}} \! \frac{U_0}{2 \mathcal{V}} c^\dagger_{\bf K}(  {\bf p}) c^\dagger_{-\bf K}({\bf q} -{\bf p})c_{{\bf K}'}({\bf k}) c_{-{\bf K}'}({\bf q} -{\bf k}) + \mathrm{h.c.}.
\end{align} 
Here, as in Eq.~\eqref{eq:H_long_wavelength}, the momenta $p, k, q < q_0$ are measured from the Weyl nodes, but we labeled the electron operators by the momenta of the Weyl nodes, $\pm {\bf K}, \pm {\bf K}'$, instead of the valley indices $a, b$. The inter-band interaction matrix element may be estimated as  $U_0 \approx V(\bm{K} - \bm{K}') - V(\bm{K} + \bm{K}')$.  
Owing to the large separation of momentum scales, $|\bm{K} - \bm{K}'| \gg k_F$, this matrix element is smaller than $V(q)$ in Eq.~\eqref{eq:H_long_wavelength} in the parameter $k_F/|{\bf K} \pm {\bf K}'|\ll 1$. To the lowest order, the effect of interband transitions is accounted for by adding  the term $\mathcal{L}^{(1)} = - \frac{1}{\mathcal{V}} \langle \hat{H}_1 \rangle $  to the Lagrangian density, where $\langle \ldots \rangle$ denotes the expectation value over the unperturbed ground state. In the mean-field approximation only the terms with ${\bf k} = {\bf p}$ contribute to this average. Denoting ${\bf q} -{\bf p} = \tilde{\bf p} $, and using  the obvious change of notations, $ab\to \pm {\bf K}$ in Eq.~\eqref{eq:anomalous_average_new},  we  get
\begin{align}
    \label{eq:H_q_expectation_value}
    \mathcal{L}^{(1)}= &   
    \int \frac{d^3 p \, d^3 \tilde{p}}{2 h^6} U_{0} 
    e^{i (\vartheta_+ +\vartheta_-)} \psi_{ {\bf K}} ({\bf p})\psi_{ -{\bf K}} (\tilde{\bf p})
   + \mathrm{h.c.} .
\end{align}
In the long wavelength limit this produces a Josephson term in the Lagrangian density,
\begin{align}
    \label{eq:L_1}
    \mathcal{L}^{(1)} = & \, J \cos (\vartheta_+ +\vartheta_-),
\end{align}
where the Josephson coupling magnitude $J$ may be estimated as
\begin{align}
    \label{eq:J_estimate}
    J \sim & \, \, \,  \lambda_{1} \Delta^2 \nu , \quad \lambda_1 = \nu U_0.
\end{align}
Here $\Delta^2 \nu$ is the energy density of exciton condensation, and $U_0$ is the inter-valley matrix element in Eq.~\eqref{eq:H_pert}.

The full Lagrangian density given by the sum of Eqs.~\eqref{eq:L_0_CDW}  and  \eqref{eq:L_1} yields the following equations of motion for the condensate phases 
\[
     \hbar^2 \nu \Ddot{\vartheta}_\alpha -  2 \alpha \bm{V}\cdot \bm{\nabla}\dot{\vartheta}_\alpha  =  \,      \frac{e^2 \bm{B}\cdot{\bm{E}}}{2\pi h c }  + \hbar^2 \rho \bm{\nabla}^2 \vartheta_\alpha - J \sin (\vartheta_+ + \vartheta_-).
\]
Introducing the symmetric and antisymmetric combinations $\vartheta_{\mathrm{s}} = \vartheta_+ + \vartheta_- $ and $\vartheta_{\mathrm{a}} = \vartheta_+ - \vartheta_- $
we get
\begin{subequations}
        \label{eq:eq_motion_sa}
\begin{align}
    \label{eq:eq_motion_s}
     \nu \hbar^2  \Ddot{\vartheta}_{\mathrm{s}}  - 2   \bm{V}\cdot \bm{\nabla}\dot{\vartheta}_{\mathrm{a}}  = & \,        \frac{ e^2 \bm{B}\cdot\bm{E}}{\pi h c }  + \rho \hbar^2  \bm{\nabla}^2 \vartheta_{\mathrm{s}} - 2 J \sin \vartheta_{\mathrm{s}}, \\
     \label{eq:eq_motion_a}
     \hbar^2 \nu \Ddot{\vartheta}_{\mathrm{a}} - 2  \bm{V}\cdot \bm{\nabla}\dot{\vartheta}_{\mathrm{s}}  = & \,   \rho \hbar^2  \bm{\nabla}^2 \vartheta_{\mathrm{a}} .
\end{align}
\end{subequations}
Note that the term $\bm{q}\cdot \bm{V} $ couples the fluctuations of $\vartheta_s$ and $\vartheta_a$ at ${\bf q} \neq 0$.

\emph{Exciton condensate modes at $\bm{B}=0$:}  At zero magnetic field, the axion coupling between $\vartheta_s$ and the electric field vanishes.  Linearizing Eqs.~\eqref{eq:eq_motion_sa} with respect to $\vartheta_s$ at $\bm{B}=0$, we get the equation that determines the spectrum of the phase fluctuations of the excitonic condensate, 
\begin{align}
    \label{eq:spectrum_zero_B}
    \omega^2_\pm = &\,\pm \sqrt{\frac{1}{4}\left( \omega_T^2 + 2 s^2 q^2 +(\bm{v}\cdot {\bf{q}})^2 \right)^2  - s^2 q^2 \left( \omega_T^2  +s^2 q^2\right) }  \nonumber \\ 
      & + \frac{1}{2} \left(\omega_T^2 + 2 s^2 q^2 +(\bm{v}\cdot {\bf{q}})^2\right)   .
\end{align}
Here we introduced the notations
\begin{align}
    \label{eq:omega_T_def}
    \omega_T^2 = &  \frac{2J}{\hbar^2 \nu}, \quad s^2=\frac{\rho}{\nu}, \quad \bm{v} = \frac{2\bm{V}}{\hbar^2 \nu}.
\end{align}

Equation~\eqref{eq:spectrum_zero_B} reproduces the well-known results for the spectrum of the collective modes in the excitonic insulators~\cite{Jerome,Halperin-Rice,Guseinov}. In the approximation of Eq.~\eqref{eq:H_long_wavelength}, which neglects the inter-node electron-electron scattering, Eq.~\eqref{eq:H_pert}, we have $J\to 0$, and the system has two gapless modes: 1) The $\omega_{-}$ mode  is the Goldstone mode associated with the spontaneous breaking of translation symmetry and corresponds to sliding of the exciton condensate,  
2) The $\omega_+$ mode is the acoustic mode that reflects conservation of the number of excitons at $J=0$.

The inter-node transitions described by Eq.~\eqref{eq:H_pert} create or annihilate pairs of excitons belonging to the different condensates. These transitions result in $J \neq 0$ and produce a gap in the spectrum of $\omega_+$. The $\omega_+$ mode is similar to the Leggett mode~\cite{Leggett} in multi-band superconductors. It is this mode that couples
to electric current at $\bm{B} \neq 0$, and corresponds to fluctuations of the axion field $\vartheta_s$. 

Sliding of the neutral exciton condensate associated with the mode $\omega_-$ does not induce charge current~\cite{Jerome}, in contrast to the sliding mode of charge density waves (CDW) in quasi-one-dimensional metals~\cite{frohlich1954theory}. This distinction between translation symmetry breaking in the excitonic insulators and in the CDW state of metals was appreciated since the early works~\cite{lee1974conductivity} on charge transport in systems with CDW order, but was lost in much of the recent literature, including the work~\cite{Grigoreva} of the present authors.

\emph{Axionic polariton at $\bm{B}\neq 0$:} Let us now consider the effect of the axion coupling \eqref{eq:axion_coupling_intro} on the electromagnetic response of WEIs at $\bm{B} \neq 0$. To this end  
we rewrite the polarization density in Eq.~\eqref{eq:Polarization_E} as 
\begin{align}
    \label{eq:polarization_theta_s}
    \bm{P} ({\bf r},t ) =    \frac{e^2}{2 \pi h c} \bm{B} \vartheta_{\mathrm{s}} ({\bf r},t ).
\end{align}
This equation shows that because of the axion coupling in Eq.~\eqref{eq:axion_coupling_intro}, at $\bm{B}\neq 0$ the deviations of $\vartheta_s$ from its equilibrium value $\bar{\vartheta}_s=0$ create electric polarization of the system along the magnetic field $\bm{B}$. Thus, at $\bm{B}\neq 0$ the phase $\vartheta_s$ becomes similar to an optical phonon displacement in an ionic crystal. In this analogy, the gap $\omega_T$ of the $\omega_+$ mode plays the role of the optical phonon frequency. Like in ionic crystals, a polariton resonance arises between the axion field $\vartheta_s$ and electromagnetic (EM) waves through the system. This affects the spectrum of EM waves and leads to magnetic birefringence. 

Since the characteristic momentum, $q \sim \omega_T/c$, of the polariton region, is small, we may evaluate the dielectric function of the system by setting $\bm{q}=0$ in Eq.~\eqref{eq:eq_motion_sa}. In this regime, the variable $\vartheta_a$  does not affect the dielectric response. Substituting the \emph{ac} electric field of the form $\bm{E}=\mathrm{Re} \left( \bm{E}_\omega e^{ - i \omega t} \right)$ into Eq.~\eqref{eq:eq_motion_sa}, neglecting the nonlinearities caused by the quantum fluctuations of $\vartheta_s$ about the equilibrium state $\vartheta_{\mathrm{s}} =0$, and using Eq.~\eqref{eq:polarization_theta_s} we get the frequency-dependent dielectric tensor $\hat{\epsilon} (\omega)$ in the form
\begin{align}
\label{eq:dielectric_tensor_q_0}
    \epsilon_{ij} (\omega) = & \epsilon^{(0)}_{ij} + \frac{2}{h^2 c^2 \nu} \left(\frac{e^2}{h}\right)^2 \frac{B_i B_j}{\omega^2_T - \omega^2}.
\end{align}
Here we used the notations of Eq.~\eqref{eq:omega_T_def} and denoted by
$\epsilon^{(0)}_{ij}$  the dielectric tensor at $\bm{B}=0$. Due to the large ratio of the exciton binding energy $\Delta$ to $\hbar \omega_T$, in the relevant frequency interval, $|\omega | \sim \omega_T  $, the first term, $\epsilon^{(0)}_{ij}$, is independent of $\omega$. Note that the axion correction to the susceptibility, second term in Eq.~\eqref{eq:dielectric_tensor_q_0}, is controlled by the gap in the exciton spectrum.

The spectrum of electromagnetic waves in WEIs is obtained by substituting the dielectric function Eq.~\eqref{eq:dielectric_tensor_q_0} into Maxwell's equations in anisotropic media~\cite{Landau8}, 
\begin{align}
    \label{eq:polariton}
\bm{q}\cdot \bm{D}_\omega = 0, \quad \omega \bm{H}_\omega= c\bm{q}\times \bm{E}_\omega
, \quad   \omega \bm{D}_\omega=- c \bm{q}\times \bm{H}_\omega.
\end{align}
The first equation here reflects the absence of external charges. This corresponds to either a vanishing or a purely transverse vector of electric induction $\bm{D}_\omega= \hat{\epsilon}_\omega \bm{E}_\omega  $. The dispersion of the longitudinal wave is given by 
\begin{align}
    \label{eq:longitudinal_wave}
    q_i\epsilon_{ij}(\omega) q_j =0.
\end{align}
For the dispersion of the two transverse waves, from Eq.~\eqref{eq:polariton} one obtains the Fresnel equation
\begin{align}
    \label{eq:Fresnel_q}
   \mathrm{det} \left( q^2 \delta_{ij} - q_i q_j -\frac{\omega^2}{c^2}\epsilon_{ij} (\omega)  \right) = 0.
\end{align}
To separate the magneto-induced anisotropy associated with the axion coupling from the crystalline anisotropy, let us consider the simplest situation, in which the dielectric function in Eq.~\eqref{eq:dielectric_tensor_q_0} is isotropic at $\bm{B}=0$, i.e. $\epsilon^{(0)}_{ij} = \epsilon (\infty) \delta_{ij}$. In this case, the dispersion depends only on the angle $\theta$ between the wavevector ${\bf q}$ and $\bm{B}$; $\cos \theta = {\bf q}\cdot \bm{B} /(qB)$.
Using Eq.~\eqref{eq:longitudinal_wave}, we obtain the frequency of the longitudinal mode,
\begin{align}
    \label{eq:omega_L}
    \omega_L^2 (\theta) = & \, \omega^2_T + \left(\frac{e^2}{h}\right)^2 \frac{2 B^2\cos^2 \theta}{\nu h^2 c^2 \epsilon(\infty)}.
\end{align}
The frequency of the longitudinal mode depends only on the direction of momentum, but not on its magnitude $|q|$, which is also the case for the plasmon frequency in WSM in the ultra-quantum limit~\cite{Spivak-Andreev}. 

The spectrum of the transverse polariton modes is determined by Eq.~\eqref{eq:Fresnel_q}. 
The ordinary wave, in which the electric field is parallel to ${\bf q}\times \bm{B}$, is unaffected by the axion coupling. The spectrum of the extraordinary wave, in which the electric field lies in the plane spanned by ${\bf q}$ and $\bm{B}$,
is given by 
\begin{align}
\label{eq:polariton_spectrum_axion}
    \omega^2_{1,2} = & \, \frac{1}{2} \left(q^2 c^2_{\infty} + \omega^2_L (0) \pm \sqrt{\left(q^2 c^2_{\infty} + \omega^2_L (0)\right)^2 - 4 q^2 c^2_{\infty} \omega^2_L (\theta)}\right).
\end{align}
Here $c^2_\infty = c^2 /\epsilon (\infty) $, and the frequencies $\omega_T$ and $\omega_L (\theta)$ are defined in Eqs.~\eqref{eq:omega_T_def} and ~\eqref{eq:omega_L}. 
The polariton spectrum in Eq.~\eqref{eq:polariton_spectrum_axion}
is similar to that in ionic crystals ~\cite{MAIANI}. In ionic crystals, $\omega_T$ corresponds to the frequency of transverse optical phonons. The frequency of the longitudinal mode increases to $\omega_L$ because of the long-range electric fields induced by the optical phonon displacements. The magnitude of this enhancement depends on the amplitude of electric polarization associated with the phonon displacement. In WEIs, the role of the optical phonon is played by oscillations of the phase $\vartheta_s$ of the exciton order parameter, and the magnitude of the associated polarization in Eq.~\eqref{eq:polarization_theta_s} arises from the axion coupling. As a result, the polariton spectrum in Eq.~\eqref{eq:polariton_spectrum_axion} is controlled by $\bm{B}$ and exhibits the uniaxial anisotropy characteristic of the chiral anomaly. 

For $\bm{E}$ parallel to $\bm{B}$, the frequency-dependent dielectric function has the same form as that for  ionic crystals, $\epsilon (\omega) = \epsilon (\infty) \frac{\omega_L^2(0) -\omega^2}{\omega_T^2 -\omega^2}$, and 
satisfies the well known Lyddane-Sachs-Teller relation,  $\epsilon_\parallel (0)/\epsilon (\infty) = \omega_L^2(0)/\omega_T^2$. Let us show that the magneto-induced enhancement factor of the static dielectric susceptibility in the excitonic phase of WSM, may become of order unity at weak magnetic fields, which satisfies our assumptions. Using Eq.~\eqref{eq:dielectric_tensor_q_0},  the enhancement factor can be expressed in terms of the cyclotron frequency $\omega_c = e v B/\hbar c k_F $ for the electrons at the Fermi level,    
\begin{align}
    \label{eq:delta_epsilon_estimate_freq}
    \frac{\epsilon(0)}{\epsilon(\infty)} -1 
     = & \, \frac{\omega_c^2}{\omega_T^2} \frac{e^2}{2\pi^2 h v}.
\end{align}
Using Eq.~\eqref{eq:omega_T_def} and the estimates \eqref{eq:J_estimate} and  we get
\begin{align}
    \label{eq:delta_epsilon_estimate}
     \frac{\epsilon(0)}{\epsilon(\infty)} -1   \sim  & \, 
    \left(\frac{\hbar \omega_c}{\Delta}\right)^2 \frac{e^2}{\hbar v \lambda_1}.
\end{align}
The assumption that the effect of the magnetic field on the exciton pairing is negligible, is valid provided the cyclotron radius exceeds the exciton coherence length $\xi \sim \hbar v/\Delta$, or equivalently $\hbar \omega_c\ll \Delta$. Since the small parameter $\hbar \omega_c/\Delta$ may be compensated by a large value of the ratio $1/\lambda_1$ in the right-hand side of Eq.~\eqref{eq:delta_epsilon_estimate} (typically, $e^2/\hbar v \sim 1$), the magneto-induced enhancement of the dielectric constant may become of order unity in the range of validity of our treatment, implying strong, $\sim 1$, axion-induced magnetic birefringence~\cite{axion_birefringence_review}  in weak magnetic fields.
This situation is similar to strong negative longitudinal magnetoresistance (NLMR) in Weyl semimetals~\cite{Son-Spivak,doi:10.1126/science.aac6089,ong2021experimental} and Dirac semimetals with a small gap~\cite{Spivak-Andreev_LNMR,doi:10.1126/sciadv.aav9771}. For Weyl/Dirac semimetals the role of the enhancement factor  $e^2/  \hbar v \lambda_1 $ is played by the large ratio of the inter-valley/helicity relaxation time to the transport relaxation time. 

In summary, we identified a new mechanism for the formation of an axionic insulator state in a minimal model of a TR-invariant compensated Weyl semimetal. Our mechanism differs from that of Ref.~\cite{wang2013chiral}, in which the gap in the electron spectrum forms exactly at the Weyl nodes  via the mechanism of chiral mass generation (see e.g. Ref.~\onlinecite{miransky2015quantum} for a review). The latter may occur only at a strong coupling and is unlikely to be realized in realistic materials because energy of the Weyl nodes rarely coincides with the Fermi level. In our mechanism, the Weyl nodes are not aligned with the Fermi level. The gap in the electron spectrum forms at the Fermi level, and is caused by the excitonic instability of the semimetal state. This may be realized at weak coupling, $e^2/\hbar v\ll 1$. As the doping decreases, the exciton gap is expected to disappear along with the Fermi energy~\cite{gor1961contribution}.
At low temperatures, the system develops two coupled exciton condensates related by TR symmetry. Although the translation symmetry is broken, the axion field is associated not with the sliding mode of the exciton condensate as proposed in Ref.~\cite{wang2013chiral}, but with the oscillations of the gapped mode that is similar to the Leggett mode~\cite{Leggett} in multiband superconductors. This has important implications for experimental searches for the axionic insulator phase in TR-invariant WSMs. 

Observation of the axionic insulator state in $(\mathrm{TaSe}_4)_2\mathrm{I}$ was recently 
reported   
in Ref.~\onlinecite{gooth2019axionic}. 
This quasi-one-dimensional material hosts 24 pairs of Weyl nodes near the Fermi energy and undergoes a transition to a CDW phase at 263K. 
The analysis of experimental data was based on the mechanism of Ref.~\onlinecite{wang2013chiral}, which encountered significant challenges~\cite{Sinchenko}. 
Our work identifies alternative experimental signatures of the axionic state -- the giant magnetic birefringence and axion-polariton resonance.

We would like to thank A. Bernevig, D. Cobden, J-H. Chu, N. Gippius, S. Kivelson, B. Spivak and V. Yudson for illuminating discussions. 
The work of A.G. and A.A. was supported by the NSF grant DMR-2424364, the Thouless Institute for Quantum Matter (TIQM), and the College of Arts and Sciences at the University of Washington. The work of L.I.G. was supported by the Office of Naval Research under the Award No. N00014-22-1-2764, and by the NSF grant DMR-2410182.


\begin{thebibliography}{41}%
	\makeatletter
	\providecommand \@ifxundefined [1]{%
		\@ifx{#1\undefined}
	}%
	\providecommand \@ifnum [1]{%
		\ifnum #1\expandafter \@firstoftwo
		\else \expandafter \@secondoftwo
		\fi
	}%
	\providecommand \@ifx [1]{%
		\ifx #1\expandafter \@firstoftwo
		\else \expandafter \@secondoftwo
		\fi
	}%
	\providecommand \natexlab [1]{#1}%
	\providecommand \enquote  [1]{``#1''}%
	\providecommand \bibnamefont  [1]{#1}%
	\providecommand \bibfnamefont [1]{#1}%
	\providecommand \citenamefont [1]{#1}%
	\providecommand \href@noop [0]{\@secondoftwo}%
	\providecommand \href [0]{\begingroup \@sanitize@url \@href}%
	\providecommand \@href[1]{\@@startlink{#1}\@@href}%
	\providecommand \@@href[1]{\endgroup#1\@@endlink}%
	\providecommand \@sanitize@url [0]{\catcode `\\12\catcode `\$12\catcode
		`\&12\catcode `\#12\catcode `\^12\catcode `\_12\catcode `\%12\relax}%
	\providecommand \@@startlink[1]{}%
	\providecommand \@@endlink[0]{}%
	\providecommand \url  [0]{\begingroup\@sanitize@url \@url }%
	\providecommand \@url [1]{\endgroup\@href {#1}{\urlprefix }}%
	\providecommand \urlprefix  [0]{URL }%
	\providecommand \Eprint [0]{\href }%
	\providecommand \doibase [0]{https://doi.org/}%
	\providecommand \selectlanguage [0]{\@gobble}%
	\providecommand \bibinfo  [0]{\@secondoftwo}%
	\providecommand \bibfield  [0]{\@secondoftwo}%
	\providecommand \translation [1]{[#1]}%
	\providecommand \BibitemOpen [0]{}%
	\providecommand \bibitemStop [0]{}%
	\providecommand \bibitemNoStop [0]{.\EOS\space}%
	\providecommand \EOS [0]{\spacefactor3000\relax}%
	\providecommand \BibitemShut  [1]{\csname bibitem#1\endcsname}%
	\let\auto@bib@innerbib\@empty
	\bibitem [{\citenamefont {Leggett}(1966)}]{Leggett}%
	\BibitemOpen
	\bibfield  {author} {\bibinfo {author} {\bibfnamefont {A.~J.}\ \bibnamefont
			{Leggett}},\ }\bibfield  {title} {\bibinfo {title} {Number-phase fluctuations
			in two-band superconductors},\ }\href {https://doi.org/10.1143/PTP.36.901}
	{\bibfield  {journal} {\bibinfo  {journal} {Progress of Theoretical Physics}\
		}\textbf {\bibinfo {volume} {36}},\ \bibinfo {pages} {901} (\bibinfo {year}
		{1966})}\BibitemShut {NoStop}%
	\bibitem [{\citenamefont {Keldysh}\ and\ \citenamefont
		{Kopaev}(1965)}]{Keldysh}%
	\BibitemOpen
	\bibfield  {author} {\bibinfo {author} {\bibfnamefont {L.~V.}\ \bibnamefont
			{Keldysh}}\ and\ \bibinfo {author} {\bibfnamefont {Y.~V.}\ \bibnamefont
			{Kopaev}},\ }\bibfield  {title} {\bibinfo {title} {Possible instability of
			the semimetallic state toward {Coulomb} interaction},\ }\href@noop {}
	{\bibfield  {journal} {\bibinfo  {journal} {Sov. Phys. Solid State}\ }\textbf
		{\bibinfo {volume} {6}},\ \bibinfo {pages} {2219} (\bibinfo {year}
		{1965})}\BibitemShut {NoStop}%
	\bibitem [{\citenamefont {Cloizeaux}(1965)}]{CLOIZEAUX}%
	\BibitemOpen
	\bibfield  {author} {\bibinfo {author} {\bibfnamefont {J.~D.}\ \bibnamefont
			{Cloizeaux}},\ }\bibfield  {title} {\bibinfo {title} {Exciton instability and
			crystallographic anomalies in semiconductors},\ }\href
	{https://doi.org/https://doi.org/10.1016/0022-3697(65)90153-8} {\bibfield
		{journal} {\bibinfo  {journal} {Journal of Physics and Chemistry of Solids}\
		}\textbf {\bibinfo {volume} {26}},\ \bibinfo {pages} {259} (\bibinfo {year}
		{1965})}\BibitemShut {NoStop}%
	\bibitem [{\citenamefont {J\'erome}\ \emph {et~al.}(1967)\citenamefont
		{J\'erome}, \citenamefont {Rice},\ and\ \citenamefont {Kohn}}]{Jerome}%
	\BibitemOpen
	\bibfield  {author} {\bibinfo {author} {\bibfnamefont {D.}~\bibnamefont
			{J\'erome}}, \bibinfo {author} {\bibfnamefont {T.~M.}\ \bibnamefont {Rice}},\
		and\ \bibinfo {author} {\bibfnamefont {W.}~\bibnamefont {Kohn}},\ }\bibfield
	{title} {\bibinfo {title} {Excitonic insulator},\ }\href
	{https://doi.org/10.1103/PhysRev.158.462} {\bibfield  {journal} {\bibinfo
			{journal} {Phys. Rev.}\ }\textbf {\bibinfo {volume} {158}},\ \bibinfo {pages}
		{462} (\bibinfo {year} {1967})}\BibitemShut {NoStop}%
	\bibitem [{\citenamefont {Keldysh}\ and\ \citenamefont
		{Kozlov}(1968)}]{Kozlov}%
	\BibitemOpen
	\bibfield  {author} {\bibinfo {author} {\bibfnamefont {L.~V.}\ \bibnamefont
			{Keldysh}}\ and\ \bibinfo {author} {\bibfnamefont {A.~N.}\ \bibnamefont
			{Kozlov}},\ }\bibfield  {title} {\bibinfo {title} {Collective properties of
			excitons in semiconductors},\ }\href@noop {} {\bibfield  {journal} {\bibinfo
			{journal} {Soviet Physics JETP}\ }\textbf {\bibinfo {volume} {27}},\ \bibinfo
		{pages} {521} (\bibinfo {year} {1968})}\BibitemShut {NoStop}%
	\bibitem [{\citenamefont {Guseinov}\ and\ \citenamefont
		{Keldysh}(1972)}]{Guseinov}%
	\BibitemOpen
	\bibfield  {author} {\bibinfo {author} {\bibfnamefont {R.}~\bibnamefont
			{Guseinov}}\ and\ \bibinfo {author} {\bibfnamefont {L.}~\bibnamefont
			{Keldysh}},\ }\bibfield  {title} {\bibinfo {title} {Nature of the phase
			transition under the conditions of an ``excitonic" instability in the
			electronic spectrum of a crystal},\ }\href@noop {} {\bibfield  {journal}
		{\bibinfo  {journal} {Soviet Physics JETP}\ }\textbf {\bibinfo {volume}
			{63}},\ \bibinfo {pages} {2255} (\bibinfo {year} {1972})}\BibitemShut
	{NoStop}%
	\bibitem [{\citenamefont {Keldysh}(2017)}]{Keldysh_rev}%
	\BibitemOpen
	\bibfield  {author} {\bibinfo {author} {\bibfnamefont {L.~V.}\ \bibnamefont
			{Keldysh}},\ }\bibfield  {title} {\bibinfo {title} {Coherent states of
			excitons},\ }\href {https://doi.org/10.3367/UFNe.2017.10.038227} {\bibfield
		{journal} {\bibinfo  {journal} {Physics-Uspekhi}\ }\textbf {\bibinfo {volume}
			{60}},\ \bibinfo {pages} {1180} (\bibinfo {year} {2017})}\BibitemShut
	{NoStop}%
	\bibitem [{\citenamefont {Li}\ and\ \citenamefont {Haldane}(2018)}]{Haldane}%
	\BibitemOpen
	\bibfield  {author} {\bibinfo {author} {\bibfnamefont {Y.}~\bibnamefont
			{Li}}\ and\ \bibinfo {author} {\bibfnamefont {F.~D.~M.}\ \bibnamefont
			{Haldane}},\ }\bibfield  {title} {\bibinfo {title} {Topological nodal
			{Cooper} pairing in doped {Weyl} metals},\ }\href
	{https://doi.org/10.1103/PhysRevLett.120.067003} {\bibfield  {journal}
		{\bibinfo  {journal} {Phys. Rev. Lett.}\ }\textbf {\bibinfo {volume} {120}},\
		\bibinfo {pages} {067003} (\bibinfo {year} {2018})}\BibitemShut {NoStop}%
	\bibitem [{\citenamefont {Grigoreva}\ \emph {et~al.}(2024)\citenamefont
		{Grigoreva}, \citenamefont {Andreev},\ and\ \citenamefont
		{Glazman}}]{Grigoreva}%
	\BibitemOpen
	\bibfield  {author} {\bibinfo {author} {\bibfnamefont {A.~A.}\ \bibnamefont
			{Grigoreva}}, \bibinfo {author} {\bibfnamefont {A.~V.}\ \bibnamefont
			{Andreev}},\ and\ \bibinfo {author} {\bibfnamefont {L.~I.}\ \bibnamefont
			{Glazman}},\ }\bibfield  {title} {\bibinfo {title} {Reconstruction of surface
			electron spectrum and cyclotron motion in the {Charge Density Wave} phase of
			{Weyl Semimetals}},\ }\href {https://doi.org/10.1103/PhysRevLett.133.166601}
	{\bibfield  {journal} {\bibinfo  {journal} {Phys. Rev. Lett.}\ }\textbf
		{\bibinfo {volume} {133}},\ \bibinfo {pages} {166601} (\bibinfo {year}
		{2024})}\BibitemShut {NoStop}%
	\bibitem [{\citenamefont {Peccei}\ and\ \citenamefont
		{Quinn}(1977)}]{peccei1977some}%
	\BibitemOpen
	\bibfield  {author} {\bibinfo {author} {\bibfnamefont {R.}~\bibnamefont
			{Peccei}}\ and\ \bibinfo {author} {\bibfnamefont {H.~R.}\ \bibnamefont
			{Quinn}},\ }\bibfield  {title} {\bibinfo {title} {Some aspects of
			instantons},\ }\href {https://doi.org/https://doi.org/10.1007/BF02730110}
	{\bibfield  {journal} {\bibinfo  {journal} {Il Nuovo Cimento A}\ }\textbf
		{\bibinfo {volume} {41}},\ \bibinfo {pages} {309} (\bibinfo {year}
		{1977})}\BibitemShut {NoStop}%
	\bibitem [{\citenamefont {Wilczek}(1987)}]{wilczek1987two}%
	\BibitemOpen
	\bibfield  {author} {\bibinfo {author} {\bibfnamefont {F.}~\bibnamefont
			{Wilczek}},\ }\bibfield  {title} {\bibinfo {title} {Two applications of axion
			electrodynamics},\ }\href {https://doi.org/10.1103/PhysRevLett.58.1799}
	{\bibfield  {journal} {\bibinfo  {journal} {Phys. Rev. Lett.}\ }\textbf
		{\bibinfo {volume} {58}},\ \bibinfo {pages} {1799} (\bibinfo {year}
		{1987})}\BibitemShut {NoStop}%
	\bibitem [{\citenamefont {Essin}\ \emph {et~al.}(2009)\citenamefont {Essin},
		\citenamefont {Moore},\ and\ \citenamefont
		{Vanderbilt}}]{essin2009magnetoelectric}%
	\BibitemOpen
	\bibfield  {author} {\bibinfo {author} {\bibfnamefont {A.~M.}\ \bibnamefont
			{Essin}}, \bibinfo {author} {\bibfnamefont {J.~E.}\ \bibnamefont {Moore}},\
		and\ \bibinfo {author} {\bibfnamefont {D.}~\bibnamefont {Vanderbilt}},\
	}\bibfield  {title} {\bibinfo {title} {Magnetoelectric polarizability and
			axion electrodynamics in crystalline insulators},\ }\href
	{https://doi.org/10.1103/PhysRevLett.102.146805} {\bibfield  {journal}
		{\bibinfo  {journal} {Phys. Rev. Lett.}\ }\textbf {\bibinfo {volume} {102}},\
		\bibinfo {pages} {146805} (\bibinfo {year} {2009})}\BibitemShut {NoStop}%
	\bibitem [{\citenamefont {Li}\ \emph {et~al.}(2010)\citenamefont {Li},
		\citenamefont {Wang}, \citenamefont {Qi},\ and\ \citenamefont
		{Zhang}}]{li2010dynamical}%
	\BibitemOpen
	\bibfield  {author} {\bibinfo {author} {\bibfnamefont {R.}~\bibnamefont
			{Li}}, \bibinfo {author} {\bibfnamefont {J.}~\bibnamefont {Wang}}, \bibinfo
		{author} {\bibfnamefont {X.-L.}\ \bibnamefont {Qi}},\ and\ \bibinfo {author}
		{\bibfnamefont {S.-C.}\ \bibnamefont {Zhang}},\ }\bibfield  {title} {\bibinfo
		{title} {Dynamical axion field in topological magnetic insulators},\ }\href
	{https://doi.org/10.1038/nphys1534} {\bibfield  {journal} {\bibinfo
			{journal} {Nature Physics}\ }\textbf {\bibinfo {volume} {6}},\ \bibinfo
		{pages} {284} (\bibinfo {year} {2010})}\BibitemShut {NoStop}%
	\bibitem [{\citenamefont {Qi}\ and\ \citenamefont
		{Zhang}(2011)}]{qi2011topological}%
	\BibitemOpen
	\bibfield  {author} {\bibinfo {author} {\bibfnamefont {X.-L.}\ \bibnamefont
			{Qi}}\ and\ \bibinfo {author} {\bibfnamefont {S.-C.}\ \bibnamefont {Zhang}},\
	}\bibfield  {title} {\bibinfo {title} {Topological insulators and
			superconductors},\ }\href {https://doi.org/10.1103/RevModPhys.83.1057}
	{\bibfield  {journal} {\bibinfo  {journal} {Rev. Mod. Phys.}\ }\textbf
		{\bibinfo {volume} {83}},\ \bibinfo {pages} {1057} (\bibinfo {year}
		{2011})}\BibitemShut {NoStop}%
	\bibitem [{\citenamefont {Wang}\ and\ \citenamefont
		{Zhang}(2013)}]{wang2013chiral}%
	\BibitemOpen
	\bibfield  {author} {\bibinfo {author} {\bibfnamefont {Z.}~\bibnamefont
			{Wang}}\ and\ \bibinfo {author} {\bibfnamefont {S.-C.}\ \bibnamefont
			{Zhang}},\ }\bibfield  {title} {\bibinfo {title} {Chiral anomaly, charge
			density waves, and axion strings from {Weyl} semimetals},\ }\href
	{https://doi.org/10.1103/PhysRevB.87.161107} {\bibfield  {journal} {\bibinfo
			{journal} {Phys. Rev. B}\ }\textbf {\bibinfo {volume} {87}},\ \bibinfo
		{pages} {161107} (\bibinfo {year} {2013})}\BibitemShut {NoStop}%
	\bibitem [{\citenamefont {Nenno}\ \emph {et~al.}(2020)\citenamefont {Nenno},
		\citenamefont {Garcia}, \citenamefont {Gooth}, \citenamefont {Felser},\ and\
		\citenamefont {Narang}}]{nenno2020axion}%
	\BibitemOpen
	\bibfield  {author} {\bibinfo {author} {\bibfnamefont {D.~M.}\ \bibnamefont
			{Nenno}}, \bibinfo {author} {\bibfnamefont {C.~A.}\ \bibnamefont {Garcia}},
		\bibinfo {author} {\bibfnamefont {J.}~\bibnamefont {Gooth}}, \bibinfo
		{author} {\bibfnamefont {C.}~\bibnamefont {Felser}},\ and\ \bibinfo {author}
		{\bibfnamefont {P.}~\bibnamefont {Narang}},\ }\bibfield  {title} {\bibinfo
		{title} {Axion physics in condensed-matter systems},\ }\href
	{https://doi.org/https://doi.org/10.1038/s42254-020-0240-2} {\bibfield
		{journal} {\bibinfo  {journal} {Nature Reviews Physics}\ }\textbf {\bibinfo
			{volume} {2}},\ \bibinfo {pages} {682} (\bibinfo {year} {2020})}\BibitemShut
	{NoStop}%
	\bibitem [{\citenamefont {Zittartz}(1968)}]{Zittartz}%
	\BibitemOpen
	\bibfield  {author} {\bibinfo {author} {\bibfnamefont {J.}~\bibnamefont
			{Zittartz}},\ }\bibfield  {title} {\bibinfo {title} {Transport properties of
			the {``Excitonic Insulator"}: Electrical conductivity},\ }\href
	{https://doi.org/10.1103/PhysRev.165.605} {\bibfield  {journal} {\bibinfo
			{journal} {Phys. Rev.}\ }\textbf {\bibinfo {volume} {165}},\ \bibinfo {pages}
		{605} (\bibinfo {year} {1968})}\BibitemShut {NoStop}%
	\bibitem [{\citenamefont {Halperin}\ and\ \citenamefont
		{Rice}(1968)}]{Halperin-Rice}%
	\BibitemOpen
	\bibfield  {author} {\bibinfo {author} {\bibfnamefont {B.~I.}\ \bibnamefont
			{Halperin}}\ and\ \bibinfo {author} {\bibfnamefont {T.~M.}\ \bibnamefont
			{Rice}},\ }\bibfield  {title} {\bibinfo {title} {Possible anomalies at a
			semimetal-semiconductor transistion},\ }\href
	{https://doi.org/10.1103/RevModPhys.40.755} {\bibfield  {journal} {\bibinfo
			{journal} {Rev. Mod. Phys.}\ }\textbf {\bibinfo {volume} {40}},\ \bibinfo
		{pages} {755} (\bibinfo {year} {1968})}\BibitemShut {NoStop}%
	\bibitem [{\citenamefont {Bell}\ and\ \citenamefont {Jackiw}(1969)}]{Bell}%
	\BibitemOpen
	\bibfield  {author} {\bibinfo {author} {\bibfnamefont {J.~S.}\ \bibnamefont
			{Bell}}\ and\ \bibinfo {author} {\bibfnamefont {R.}~\bibnamefont {Jackiw}},\
	}\bibfield  {title} {\bibinfo {title} {A {PCAC} puzzle: {$\pi^{0} \rightarrow
				\gamma \gamma$} in the {$\sigma$} - model},\ }\href
	{https://doi.org/https://doi.org/10.1007/BF02823296} {\bibfield  {journal}
		{\bibinfo  {journal} {Il Nuovo Cimento A}\ }\textbf {\bibinfo {volume}
			{60}},\ \bibinfo {pages} {47} (\bibinfo {year} {1969})}\BibitemShut {NoStop}%
	\bibitem [{\citenamefont {Adler}(1969)}]{Adler}%
	\BibitemOpen
	\bibfield  {author} {\bibinfo {author} {\bibfnamefont {S.~L.}\ \bibnamefont
			{Adler}},\ }\bibfield  {title} {\bibinfo {title} {Axial-vector vertex in
			spinor electrodynamics},\ }\href {https://doi.org/10.1103/PhysRev.177.2426}
	{\bibfield  {journal} {\bibinfo  {journal} {Phys. Rev.}\ }\textbf {\bibinfo
			{volume} {177}},\ \bibinfo {pages} {2426} (\bibinfo {year}
		{1969})}\BibitemShut {NoStop}%
	\bibitem [{\citenamefont {Vilenkin}(1980)}]{Vilenkin}%
	\BibitemOpen
	\bibfield  {author} {\bibinfo {author} {\bibfnamefont {A.}~\bibnamefont
			{Vilenkin}},\ }\bibfield  {title} {\bibinfo {title} {Equilibrium
			parity-violating current in a magnetic field},\ }\href
	{https://doi.org/10.1103/PhysRevD.22.3080} {\bibfield  {journal} {\bibinfo
			{journal} {Phys. Rev. D}\ }\textbf {\bibinfo {volume} {22}},\ \bibinfo
		{pages} {3080} (\bibinfo {year} {1980})}\BibitemShut {NoStop}%
	\bibitem [{\citenamefont {Nielsen}\ and\ \citenamefont
		{Ninomiya}(1983)}]{NIELSEN_NLMR}%
	\BibitemOpen
	\bibfield  {author} {\bibinfo {author} {\bibfnamefont {H.}~\bibnamefont
			{Nielsen}}\ and\ \bibinfo {author} {\bibfnamefont {M.}~\bibnamefont
			{Ninomiya}},\ }\bibfield  {title} {\bibinfo {title} {The {Adler-Bell-Jackiw}
			anomaly and {Weyl} fermions in a crystal},\ }\href
	{https://doi.org/https://doi.org/10.1016/0370-2693(83)91529-0} {\bibfield
		{journal} {\bibinfo  {journal} {Physics Letters B}\ }\textbf {\bibinfo
			{volume} {130}},\ \bibinfo {pages} {389} (\bibinfo {year}
		{1983})}\BibitemShut {NoStop}%
	\bibitem [{Note1()}]{Note1}%
	\BibitemOpen
	\bibinfo {note} {This rate can be understood as follows~{\protect \color
			{magenta}\cite {NIELSEN_NLMR}}. The spectral flow arises from the zeroth
		Landau band, which has a chiral spectrum $\epsilon (p_z) = k_a v p_z$, where
		$p_z$ is the momentum along $\protect \mbox {\protect \boldmath {$B$}}$. The
		acceleration of the electrons by the electric field, $\protect \dot {p}_z = e
		E_z $, creates the flow of electron levels (spectral flow) between the
		valence and conduction bands. The spectral flow rate is given by the product
		of the Landau level degeneracy $e B/{hc}$ and $ \protect \dot
		{p}_z/h$.}\BibitemShut {Stop}%
	\bibitem [{Note2()}]{Note2}%
	\BibitemOpen
	\bibinfo {note} {In a finite system, this would produce a charge buildup at
		the boundary, creating macroscopic electric fields and longitudinal
		oscillations, see also Eq.~\protect \textup {\hbox {\mathsurround \z@
				\protect \normalfont (\ignorespaces \ref {eq:omega_L}\unskip \@@italiccorr
				)}} below.}\BibitemShut {Stop}%
	\bibitem [{Note3()}]{Note3}%
	\BibitemOpen
	\bibinfo {note} {This is a condition that the exciton condensation opens a
		spectral gap at the local chemical potential. It applies not only in
		equilibrium but also in the presence of external perturbations with
		frequencies $\hbar \omega \ll \Delta $}\BibitemShut {NoStop}%
	\bibitem [{\citenamefont {Landau}\ and\ \citenamefont
		{Lifshitz}(1980)}]{Landau2}%
	\BibitemOpen
	\bibfield  {author} {\bibinfo {author} {\bibfnamefont {L.~D.}\ \bibnamefont
			{Landau}}\ and\ \bibinfo {author} {\bibfnamefont {E.~M.}\ \bibnamefont
			{Lifshitz}},\ }\href@noop {} {\emph {\bibinfo {title} {The classical theory
				of fields}}}\ (\bibinfo  {publisher} {Butterworth-Heinemann},\ \bibinfo
	{year} {1980})\BibitemShut {NoStop}%
	\bibitem [{\citenamefont {Fr{\"o}hlich}(1954)}]{frohlich1954theory}%
	\BibitemOpen
	\bibfield  {author} {\bibinfo {author} {\bibfnamefont {H.}~\bibnamefont
			{Fr{\"o}hlich}},\ }\bibfield  {title} {\bibinfo {title} {On the theory of
			superconductivity: the one-dimensional case},\ }\href
	{https://doi.org/https://doi.org/10.1098/rspa.1954.0116} {\bibfield
		{journal} {\bibinfo  {journal} {Proceedings of the Royal Society of London.
				Series A. Mathematical and Physical Sciences}\ }\textbf {\bibinfo {volume}
			{223}},\ \bibinfo {pages} {296} (\bibinfo {year} {1954})}\BibitemShut
	{NoStop}%
	\bibitem [{\citenamefont {Lee}\ \emph {et~al.}(1974)\citenamefont {Lee},
		\citenamefont {Rice},\ and\ \citenamefont {Anderson}}]{lee1974conductivity}%
	\BibitemOpen
	\bibfield  {author} {\bibinfo {author} {\bibfnamefont {P.}~\bibnamefont
			{Lee}}, \bibinfo {author} {\bibfnamefont {T.}~\bibnamefont {Rice}},\ and\
		\bibinfo {author} {\bibfnamefont {P.}~\bibnamefont {Anderson}},\ }\bibfield
	{title} {\bibinfo {title} {Conductivity from charge or spin density waves},\
	}\href {https://doi.org/https://doi.org/10.1016/0038-1098(74)90868-0}
	{\bibfield  {journal} {\bibinfo  {journal} {Solid State Communications}\
		}\textbf {\bibinfo {volume} {14}},\ \bibinfo {pages} {703} (\bibinfo {year}
		{1974})}\BibitemShut {NoStop}%
	\bibitem [{\citenamefont {Landau}\ and\ \citenamefont
		{Lifshitz}(1960)}]{Landau8}%
	\BibitemOpen
	\bibfield  {author} {\bibinfo {author} {\bibfnamefont {L.~D.}\ \bibnamefont
			{Landau}}\ and\ \bibinfo {author} {\bibfnamefont {E.~M.}\ \bibnamefont
			{Lifshitz}},\ }\href@noop {} {\emph {\bibinfo {title} {Electrodynamics of
				Continuous Media}}}\ (\bibinfo  {publisher} {Pergamon Press},\ \bibinfo
	{year} {1960})\BibitemShut {NoStop}%
	\bibitem [{\citenamefont {Spivak}\ and\ \citenamefont
		{Andreev}(2016)}]{Spivak-Andreev}%
	\BibitemOpen
	\bibfield  {author} {\bibinfo {author} {\bibfnamefont {B.~Z.}\ \bibnamefont
			{Spivak}}\ and\ \bibinfo {author} {\bibfnamefont {A.~V.}\ \bibnamefont
			{Andreev}},\ }\bibfield  {title} {\bibinfo {title} {Magnetotransport
			phenomena related to the chiral anomaly in {Weyl} semimetals},\ }\href
	{https://doi.org/10.1103/PhysRevB.93.085107} {\bibfield  {journal} {\bibinfo
			{journal} {Phys. Rev. B}\ }\textbf {\bibinfo {volume} {93}},\ \bibinfo
		{pages} {085107} (\bibinfo {year} {2016})}\BibitemShut {NoStop}%
	\bibitem [{\citenamefont {Maiani}\ \emph {et~al.}(1986)\citenamefont {Maiani},
		\citenamefont {Petronzio},\ and\ \citenamefont {Zavattini}}]{MAIANI}%
	\BibitemOpen
	\bibfield  {author} {\bibinfo {author} {\bibfnamefont {L.}~\bibnamefont
			{Maiani}}, \bibinfo {author} {\bibfnamefont {R.}~\bibnamefont {Petronzio}},\
		and\ \bibinfo {author} {\bibfnamefont {E.}~\bibnamefont {Zavattini}},\
	}\bibfield  {title} {\bibinfo {title} {Effects of nearly massless, spin-zero
			particles on light propagation in a magnetic field},\ }\href
	{https://doi.org/https://doi.org/10.1016/0370-2693(86)90869-5} {\bibfield
		{journal} {\bibinfo  {journal} {Physics Letters B}\ }\textbf {\bibinfo
			{volume} {175}},\ \bibinfo {pages} {359} (\bibinfo {year}
		{1986})}\BibitemShut {NoStop}%
	\bibitem [{\citenamefont {Ejlli}\ \emph {et~al.}(2020)\citenamefont {Ejlli},
		\citenamefont {{Della Valle}}, \citenamefont {Gastaldi}, \citenamefont
		{Messineo}, \citenamefont {Pengo}, \citenamefont {Ruoso},\ and\ \citenamefont
		{Zavattini}}]{axion_birefringence_review}%
	\BibitemOpen
	\bibfield  {author} {\bibinfo {author} {\bibfnamefont {A.}~\bibnamefont
			{Ejlli}}, \bibinfo {author} {\bibfnamefont {F.}~\bibnamefont {{Della
					Valle}}}, \bibinfo {author} {\bibfnamefont {U.}~\bibnamefont {Gastaldi}},
		\bibinfo {author} {\bibfnamefont {G.}~\bibnamefont {Messineo}}, \bibinfo
		{author} {\bibfnamefont {R.}~\bibnamefont {Pengo}}, \bibinfo {author}
		{\bibfnamefont {G.}~\bibnamefont {Ruoso}},\ and\ \bibinfo {author}
		{\bibfnamefont {G.}~\bibnamefont {Zavattini}},\ }\bibfield  {title} {\bibinfo
		{title} {The {PVLAS} experiment: A 25 year effort to measure vacuum magnetic
			birefringence},\ }\href
	{https://doi.org/https://doi.org/10.1016/j.physrep.2020.06.001} {\bibfield
		{journal} {\bibinfo  {journal} {Physics Reports}\ }\textbf {\bibinfo {volume}
			{871}},\ \bibinfo {pages} {1} (\bibinfo {year} {2020})}\BibitemShut {NoStop}%
	\bibitem [{\citenamefont {Son}\ and\ \citenamefont
		{Spivak}(2013)}]{Son-Spivak}%
	\BibitemOpen
	\bibfield  {author} {\bibinfo {author} {\bibfnamefont {D.~T.}\ \bibnamefont
			{Son}}\ and\ \bibinfo {author} {\bibfnamefont {B.~Z.}\ \bibnamefont
			{Spivak}},\ }\bibfield  {title} {\bibinfo {title} {Chiral anomaly and
			classical negative magnetoresistance of {Weyl} metals},\ }\href
	{https://doi.org/10.1103/PhysRevB.88.104412} {\bibfield  {journal} {\bibinfo
			{journal} {Phys. Rev. B}\ }\textbf {\bibinfo {volume} {88}},\ \bibinfo
		{pages} {104412} (\bibinfo {year} {2013})}\BibitemShut {NoStop}%
	\bibitem [{\citenamefont {Xiong}\ \emph {et~al.}(2015)\citenamefont {Xiong},
		\citenamefont {Kushwaha}, \citenamefont {Liang}, \citenamefont {Krizan},
		\citenamefont {Hirschberger}, \citenamefont {Wang}, \citenamefont {Cava},\
		and\ \citenamefont {Ong}}]{doi:10.1126/science.aac6089}%
	\BibitemOpen
	\bibfield  {author} {\bibinfo {author} {\bibfnamefont {J.}~\bibnamefont
			{Xiong}}, \bibinfo {author} {\bibfnamefont {S.~K.}\ \bibnamefont {Kushwaha}},
		\bibinfo {author} {\bibfnamefont {T.}~\bibnamefont {Liang}}, \bibinfo
		{author} {\bibfnamefont {J.~W.}\ \bibnamefont {Krizan}}, \bibinfo {author}
		{\bibfnamefont {M.}~\bibnamefont {Hirschberger}}, \bibinfo {author}
		{\bibfnamefont {W.}~\bibnamefont {Wang}}, \bibinfo {author} {\bibfnamefont
			{R.~J.}\ \bibnamefont {Cava}},\ and\ \bibinfo {author} {\bibfnamefont
			{N.~P.}\ \bibnamefont {Ong}},\ }\bibfield  {title} {\bibinfo {title}
		{Evidence for the chiral anomaly in the {Dirac} semimetal {Na$_3$Bi}},\
	}\href {https://doi.org/10.1126/science.aac6089} {\bibfield  {journal}
		{\bibinfo  {journal} {Science}\ }\textbf {\bibinfo {volume} {350}},\ \bibinfo
		{pages} {413} (\bibinfo {year} {2015})}\BibitemShut {NoStop}%
	\bibitem [{\citenamefont {Ong}\ and\ \citenamefont
		{Liang}(2021)}]{ong2021experimental}%
	\BibitemOpen
	\bibfield  {author} {\bibinfo {author} {\bibfnamefont {N.}~\bibnamefont
			{Ong}}\ and\ \bibinfo {author} {\bibfnamefont {S.}~\bibnamefont {Liang}},\
	}\bibfield  {title} {\bibinfo {title} {Experimental signatures of the chiral
			anomaly in {Dirac--Weyl} semimetals},\ }\href
	{https://doi.org/https://doi.org/10.1038/s42254-021-00310-9} {\bibfield
		{journal} {\bibinfo  {journal} {Nature Reviews Physics}\ }\textbf {\bibinfo
			{volume} {3}},\ \bibinfo {pages} {394} (\bibinfo {year} {2021})}\BibitemShut
	{NoStop}%
	\bibitem [{\citenamefont {Andreev}\ and\ \citenamefont
		{Spivak}(2018)}]{Spivak-Andreev_LNMR}%
	\BibitemOpen
	\bibfield  {author} {\bibinfo {author} {\bibfnamefont {A.~V.}\ \bibnamefont
			{Andreev}}\ and\ \bibinfo {author} {\bibfnamefont {B.~Z.}\ \bibnamefont
			{Spivak}},\ }\bibfield  {title} {\bibinfo {title} {Longitudinal negative
			magnetoresistance and magnetotransport phenomena in conventional and
			topological conductors},\ }\href
	{https://doi.org/10.1103/PhysRevLett.120.026601} {\bibfield  {journal}
		{\bibinfo  {journal} {Phys. Rev. Lett.}\ }\textbf {\bibinfo {volume} {120}},\
		\bibinfo {pages} {026601} (\bibinfo {year} {2018})}\BibitemShut {NoStop}%
	\bibitem [{\citenamefont {Mutch}\ \emph {et~al.}(2019)\citenamefont {Mutch},
		\citenamefont {Chen}, \citenamefont {Went}, \citenamefont {Qian},
		\citenamefont {Wilson}, \citenamefont {Andreev}, \citenamefont {Chen},\ and\
		\citenamefont {Chu}}]{doi:10.1126/sciadv.aav9771}%
	\BibitemOpen
	\bibfield  {author} {\bibinfo {author} {\bibfnamefont {J.}~\bibnamefont
			{Mutch}}, \bibinfo {author} {\bibfnamefont {W.-C.}\ \bibnamefont {Chen}},
		\bibinfo {author} {\bibfnamefont {P.}~\bibnamefont {Went}}, \bibinfo {author}
		{\bibfnamefont {T.}~\bibnamefont {Qian}}, \bibinfo {author} {\bibfnamefont
			{I.~Z.}\ \bibnamefont {Wilson}}, \bibinfo {author} {\bibfnamefont
			{A.}~\bibnamefont {Andreev}}, \bibinfo {author} {\bibfnamefont {C.-C.}\
			\bibnamefont {Chen}},\ and\ \bibinfo {author} {\bibfnamefont {J.-H.}\
			\bibnamefont {Chu}},\ }\bibfield  {title} {\bibinfo {title} {Evidence for a
			strain-tuned topological phase transition in {ZrTe$_5$}},\ }\href
	{https://doi.org/10.1126/sciadv.aav9771} {\bibfield  {journal} {\bibinfo
			{journal} {Science Advances}\ }\textbf {\bibinfo {volume} {5}},\ \bibinfo
		{pages} {eaav9771} (\bibinfo {year} {2019})}\BibitemShut {NoStop}%
	\bibitem [{\citenamefont {Miransky}\ and\ \citenamefont
		{Shovkovy}(2015)}]{miransky2015quantum}%
	\BibitemOpen
	\bibfield  {author} {\bibinfo {author} {\bibfnamefont {V.~A.}\ \bibnamefont
			{Miransky}}\ and\ \bibinfo {author} {\bibfnamefont {I.~A.}\ \bibnamefont
			{Shovkovy}},\ }\bibfield  {title} {\bibinfo {title} {Quantum field theory in
			a magnetic field: From quantum chromodynamics to graphene and {Dirac}
			semimetals},\ }\href
	{https://doi.org/https://doi.org/10.1016/j.physrep.2015.02.003} {\bibfield
		{journal} {\bibinfo  {journal} {Physics Reports}\ }\textbf {\bibinfo {volume}
			{576}},\ \bibinfo {pages} {1} (\bibinfo {year} {2015})}\BibitemShut {NoStop}%
	\bibitem [{\citenamefont {Gor’kov}\ and\ \citenamefont
		{Melik-Barkhudarov}(1961)}]{gor1961contribution}%
	\BibitemOpen
	\bibfield  {author} {\bibinfo {author} {\bibfnamefont {L.}~\bibnamefont
			{Gor’kov}}\ and\ \bibinfo {author} {\bibfnamefont {T.}~\bibnamefont
			{Melik-Barkhudarov}},\ }\bibfield  {title} {\bibinfo {title} {Contribution to
			the theory of superfluidity in an imperfect {Fermi} gas},\ }\href@noop {}
	{\bibfield  {journal} {\bibinfo  {journal} {Soviet Physics JETP}\ }\textbf
		{\bibinfo {volume} {13}},\ \bibinfo {pages} {1018} (\bibinfo {year}
		{1961})}\BibitemShut {NoStop}%
	\bibitem [{\citenamefont {Gooth}\ \emph {et~al.}(2019)\citenamefont {Gooth},
		\citenamefont {Bradlyn}, \citenamefont {Honnali}, \citenamefont {Schindler},
		\citenamefont {Kumar}, \citenamefont {Noky}, \citenamefont {Qi},
		\citenamefont {Shekhar}, \citenamefont {Sun}, \citenamefont {Wang} \emph
		{et~al.}}]{gooth2019axionic}%
	\BibitemOpen
	\bibfield  {author} {\bibinfo {author} {\bibfnamefont {J.}~\bibnamefont
			{Gooth}}, \bibinfo {author} {\bibfnamefont {B.}~\bibnamefont {Bradlyn}},
		\bibinfo {author} {\bibfnamefont {S.}~\bibnamefont {Honnali}}, \bibinfo
		{author} {\bibfnamefont {C.}~\bibnamefont {Schindler}}, \bibinfo {author}
		{\bibfnamefont {N.}~\bibnamefont {Kumar}}, \bibinfo {author} {\bibfnamefont
			{J.}~\bibnamefont {Noky}}, \bibinfo {author} {\bibfnamefont {Y.}~\bibnamefont
			{Qi}}, \bibinfo {author} {\bibfnamefont {C.}~\bibnamefont {Shekhar}},
		\bibinfo {author} {\bibfnamefont {Y.}~\bibnamefont {Sun}}, \bibinfo {author}
		{\bibfnamefont {Z.}~\bibnamefont {Wang}}, \emph {et~al.},\ }\bibfield
	{title} {\bibinfo {title} {Axionic {Charge Density Wave} in the {Weyl}
			semimetal {(TaSe$_{4}$)$_{2}$I}},\ }\href
	{https://doi.org/https://doi.org/10.1038/s41586-019-1630-4} {\bibfield
		{journal} {\bibinfo  {journal} {Nature}\ }\textbf {\bibinfo {volume} {575}},\
		\bibinfo {pages} {315} (\bibinfo {year} {2019})}\BibitemShut {NoStop}%
	\bibitem [{\citenamefont {Sinchenko}\ \emph {et~al.}(2022)\citenamefont
		{Sinchenko}, \citenamefont {Ballou}, \citenamefont {Lorenzo}, \citenamefont
		{Grenet},\ and\ \citenamefont {Monceau}}]{Sinchenko}%
	\BibitemOpen
	\bibfield  {author} {\bibinfo {author} {\bibfnamefont {A.~A.}\ \bibnamefont
			{Sinchenko}}, \bibinfo {author} {\bibfnamefont {R.}~\bibnamefont {Ballou}},
		\bibinfo {author} {\bibfnamefont {J.~E.}\ \bibnamefont {Lorenzo}}, \bibinfo
		{author} {\bibfnamefont {T.}~\bibnamefont {Grenet}},\ and\ \bibinfo {author}
		{\bibfnamefont {P.}~\bibnamefont {Monceau}},\ }\bibfield  {title} {\bibinfo
		{title} {Does {$(\mathrm{TaSe}_{4})_{2}\mathrm{I}$} really harbor an axionic
			charge density wave?},\ }\href {https://doi.org/10.1063/5.0080380} {\bibfield
		{journal} {\bibinfo  {journal} {Applied Physics Letters}\ }\textbf {\bibinfo
			{volume} {120}},\ \bibinfo {pages} {063102} (\bibinfo {year}
		{2022})}\BibitemShut {NoStop}%
\end{thebibliography}

%

\end{document}